\documentclass[10pt]{article}
\begin{document}
\begin{center}
{\Large{\bf Reply to Comment on Dirac spectral sum rules for QCD 
in three dimensions
}}\\[2mm]

{
U.~Magnea \\
{\small INFN, Via P. Giuria 1, I-10125 Torino, Italy \\
blom@to.infn.it}
}\\[2mm]

{\small I reply to the comment by Dr S. Nishigaki (hep-th/0007042)
to my papers Phys. Rev. D61 (2000) 056005 and Phys. Rev. D62 (2000) 016005.
}
\end{center}

Recently, Nishigaki 
\cite{Comment} pointed out that the first sum rule for the
eigenvalues of the Dirac operator given in my papers \cite{QCD3_1,QCD3_2} 
was incorrect. This appears to be true, and I thank 
Dr Nishigaki for drawing attention to this fact. That the sum rules in 
\cite{QCD3_1,QCD3_2} do not agree with the results by Hilmoine and Niclasen 
\cite{ChrRu} had been communicated to me by these authors
already before the appearance of their paper, but I was at the time 
unable to detect where my calculation failed.

However, I do not agree with the explantion by Dr Nishigaki of why the
sum rules were not reproduced correctly.
According to him, the authors of \cite{V25}
``were aware that they needed to extend the saddle point manifold... 
in order to obtain a correct result''. Such an extension of the integral
over the coset obviously can not have any effect on the result, since
this procedure only involves integrating over dummy variables on which the
integrand does not depend (namely the degrees of freedom of the stability 
subgroup), and amounts only to multiplying the integral by a numerical
factor. This extension was done in \cite{V25} only for convenience. 
Also, to my understanding, Dr Nishigaki's references to the papers by Kamenev
and M{\`e}zard are not relevant in this context.

Instead, the reason is quite simple. In \cite{QCD3_1}, 
formula (6.47),
and in \cite{QCD3_2}, formula (8.60), in order to obtain the right sum rule, 
the sum should go from $1$ to $M_{as}$ and from $1$ to $M_{s}$ respectively,
where $M_{as}$ is the number of antisymmetric traceless generators that
generate antisymmetric unimodular matrices, and $M_s$ is the
number of symmetric traceless generators that generate symmetric 
unimodular matrices. Taking into account the sizes of these 
matrices in the two cases we get

\begin{equation}
M_{as}=\frac{4N_f(4N_f-1)}{2} - 1, \ \ \ \ \ \ M_s=\frac{2N_f(2N_f+1)}{2} - 1
\end{equation}

We then obtain for $\beta =1$ 

\begin{equation}
\left\langle \sum_{\lambda_k > 0} 
\frac{1}{(N\Sigma \lambda_k)^2} \right\rangle = \frac{2N_f}{2(2N_f-1)(4N_f+1)}
\end{equation}

and for $\beta =4$ 

\begin{equation}
\left\langle \sum_{\lambda_k > 0} 
\frac{1}{(N\Sigma \lambda_k)^2} \right\rangle = \frac{2N_f}{2(2N_f-1)(N_f+1)}
\end{equation}

This is in total agreement with the sum rules obtained by Hilmoine and 
Niclasen and by Nishigaki. I remind the reader that my definition of
$N_f$ differs by a factor of two from the other authors' definition, and that 
I sum only over the positive eigenvalues.
In the sum rule for $\beta =1$, in addition to the above mentioned 
correction, the sum rule was off by a factor of 4 due to various mistakes
of factors of 2 for which I apologize. I will
update the electronic versions of \cite{QCD3_1,QCD3_2} shortly.

The $3d$ sum rules can be summarized as follows:

\begin{equation}
\left\langle \sum_{\lambda_k > 0} 
\frac{1}{(N\Sigma \lambda_k)^2} \right\rangle = \frac{2N_f}{2(2N_f-1)
(\frac{4N_f}{\beta }+1)}
\end{equation}

where the number of flavors is always even and denoted by $2N_f$, and 
$\beta $ is the Dyson index of the corresponding matrix model.
The sum rule for $\beta =2$ was given by Verbaarschot and Zahed in \cite{V25}.
\vskip5mm

{\bf Acknowledgment}

I thank Jac Verbaarschot for correspondence. I am indebted to him for
assisting me in finding the mistake in my sum rules \cite{Jac}.

\end{document}